\documentclass[namedreferences,optionalrh,hyperref]{spr-sola}

\usepackage{amssymb}
\usepackage[utf8]{inputenc}

\usepackage[utf8]{inputenc}
\usepackage{graphicx}
\graphicspath{{figures/}}
\usepackage{tabularx}
\usepackage{newtxtext}

\usepackage{color}
\usepackage[shortlabels]{enumitem}
\usepackage{subfig} 
\usepackage{datetime}
\usepackage{cancel}
\usepackage[normalem]{ulem}

\usepackage{pifont}

\newcolumntype{C}[1]{>{\centering\arraybackslash}p{#1}}
\newcolumntype{L}[1]{>{\raggedright\arraybackslash}p{#1}}
\newcolumntype{R}[1]{>{\raggedleft\let\newline\\\arraybackslash\hspace{0pt}}m{#1}}
\newcolumntype{^}{>{\currentrowstyle}}

\newcommand{\cmdint}[1]{%
  {\scantokens{#1\noexpand}}%
  \endgroup
}

\usepackage[nolist]{acronym}

\definecolor{mpg}{rgb}{0.,0.48627451,0.4745098}
\definecolor{mpgdark}{rgb}{0.,0.36470588, 0.35588235}
\definecolor{linka}{rgb}{0.0,0.0,0.7}
\definecolor{linkb}{rgb}{0.0,0.0,0.5}
\definecolor{linkc}{rgb}{0.0,0.0,0.3}
\definecolor{cmt}{rgb}{0.5,0.0,0.0}
\definecolor{todo}{rgb}{0.6,0.4,0.8}
\definecolor{shade}{rgb}{0.7,0.7,0.7}
\definecolor{akl}{rgb}{0.6,0.2,0.0}
\definecolor{jcdti}{rgb}{0.6,0.2,0.6}
\definecolor{sn}{rgb}{0.8,0.2,0.6}
\definecolor{sks}{rgb}{0.8,0.1,0.0}

\newcommand{\akl}[1]{{\color{akl}\textit{!!! Andi: [#1]}}}

\newcommand{\sunrise}{\textsc{Sunrise}}
\newcommand{\sunrisei}{\textsc{Sunrise~i}}
\newcommand{\sunriseiandii}{\textsc{Sunrise~i} and \textsc{ii}}
\newcommand{\sunriseii}{\textsc{Sunrise~ii}}
\newcommand{\sunriseiii}{\textsc{Sunrise~iii}}
\newcommand{\hinode}{\textit{Hinode}}
\newcommand{\halpha}{H$\alpha$}

\newcommand{\spc}{\hbox{S$^3$PC}}

\begin{document}

    
\begin{opening}



\title{\sunriseiii{}: The Wavefront Correction System}


%
%
%
\author[addressref={kis},email={thomas.berkefeld@leibniz-kis.de}]{\inits{T.}\fnm{Thomas}~\lnm{Berkefeld}}
\author[addressref={kis}]{\inits{A.}~\fnm{Alexander}~\lnm{Bell}}
\author[addressref={kis}]{\inits{R.}~\fnm{Reiner}~\lnm{Volkmer}}
\author[addressref={kis}]{\inits{F.}~\fnm{Frank}~\lnm{Heidecke}}
\author[addressref={kis}]{\inits{T.}~\fnm{Tobias}~\lnm{Preis}}
\author[addressref={kis}]{\inits{T.}~\fnm{Thomas}~\lnm{Sonner}}
\author[addressref={kis}]{\inits{E.}~\fnm{Eiji}~\lnm{Nakai}}
\author[addressref={mps},corref,email={lagg@mps.mpg.de}]{\inits{A.}~\fnm{Andreas}~\lnm{Korpi-Lagg}\orcid{0000-0003-1459-7074}}
\author[addressref={mps},email={gandorfer@mps.mpg.de}]{\inits{A.}~\fnm{Achim}~\lnm{Gandorfer}\orcid{0000-0002-9972-9840}}
\author[addressref={mps},email={solanki@mps.mpg.de}]{\inits{S.~K.}~\fnm{Sami~K.}~\lnm{Solanki}\orcid{0000-0002-3418-8449}}
\author[addressref={iaa,S3PC},email={jti@iaa.es}]{\inits{J.~C.}~\fnm{Jose~Carlos}~\lnm{del~Toro~Iniesta}\orcid{0000-0002-3387-026X}}
\author[addressref={naoj,unitokyo,grad},email={yukio.katsukawa@nao.ac.jp}]{\inits{Y.}~\fnm{Yukio}~\lnm{Katsukawa}\orcid{0000-0002-5054-8782}}
\author[addressref={apl},email={pietro.bernasconi@jhuapl.edu}]{\inits{P.}~\fnm{Pietro}~\lnm{Bernasconi}\orcid{0000-0002-0787-8954}}
\author[addressref={mps},email={feller@mps.mpg.de}]{\inits{A.}~\fnm{Alex}~\lnm{Feller}\orcid{0009-0009-4425-599X}}
\author[addressref={mps},email={riethmueller@mps.mpg.de}]{\inits{T.~L.}~\fnm{Tino~L.}~\lnm{Riethmüller}\orcid{0000-0001-6317-4380}}
\author[addressref={inta,S3PC}]{\inits{A.}\fnm{Alberto}~\lnm{Álvarez-Herrero}\orcid{0000-0001-9228-3412}}
\author[addressref={naoj}]{\inits{M.}\fnm{Masahito}~\lnm{Kubo}\orcid{0000-0001-5616-2808}}
\author[addressref={iac}]{\inits{V.}\fnm{Valentín}~\lnm{Martínez~Pillet}\orcid{0000-0001-7764-6895}}
\author[addressref={mps}]{\inits{S.}\fnm{H.~N.}~\lnm{Smitha}\orcid{0000-0003-3490-6532}}
\author[addressref={iaa,S3PC}]{\inits{D.}\fnm{David}~\lnm{Orozco~Suárez}\orcid{0000-0001-8829-1938}}
\author[addressref={mps}]{\inits{B.}\fnm{Bianca}~\lnm{Grauf}}
\author[addressref={apl}]{\inits{M.}\fnm{Michael}~\lnm{Carpenter}}

\address[id={kis}]{Institut für Sonnenphysik (KIS), Georges-Köhler-Allee 401a, 79110 Freiburg, Germany}
\address[id={mps}]{Max-Planck-Institut für Sonnensystemforschung, Justus-von-Liebig-Weg 3, 37077 Göttingen, Germany}
\address[id={iaa}]{Instituto de Astrofísica de Andalucía, CSIC, Glorieta de la Astronomía s/n, 18008 Granada, Spain}
\address[id={S3PC}]{Spanish Space Solar Physics Consortium (\href{https://s3pc.es}{\spc})}
\address[id={naoj}]{National Astronomical Observatory of Japan, 2-21-1 Osawa, Mitaka, Tokyo 181-8588, Japan}
\address[id={unitokyo}]{Department of Astronomy, The University of Tokyo, 7-3-1, Hongo, Bunkyo-ku, Tokyo 113-0033, Japan}
\address[id={grad}]{The Graduate University for Advanced Studies, Sokendai, Shonan Village, Hayama, Kanagawa 240-0193, Japan}
\address[id={apl}]{Johns Hopkins University Applied Physics Laboratory, 11100 Johns Hopkins Road, Laurel, MD, USA}
\address[id={inta}]{Instituto Nacional de T\'ecnica Aeroespacial (INTA), Ctra. de Ajalvir, km. 4, E-28850 Torrejón de Ardoz, Spain}
\address[id={iac}]{Instituto de Astrof\'{\i}sica de Canarias, V\'{\i}a L\'actea, s/n, E-38205 La Laguna, Spain}

\date{Received \today; \currenttime}

\newpage


 
\begin{abstract}
This paper describes the wave-front correction and image stabilisation system (CWS) developed for
the \sunriseiii{} balloon-borne telescope, and provides information about its performance as measured during the integration into the telescope and during the 2024 science flight.
The fast image stabilisation is done by a correlation tracker (CT) and a fast tip-tilt mirror, 
low order aberrations such as defocus and coma are measured by a six-element Shack-Hartmann wavefront
sensor (WFS) and corrected by an active telescope secondary mirror for automated focus and manual coma correction. 
The CWS is specified to deliver a stabilised image with a precision of 0.005 arcsec (rms).
The autofocus adjustment is specified to maintain a focus stability of 0.01 waves in the focal plane of the
CWS. 
\end{abstract}



\end{opening}



\section{Introduction}
\label{sec:intro} 

\begin{figure} [ht]
   \begin{center}
   \begin{tabular}{c} 
   \includegraphics[height=7cm]{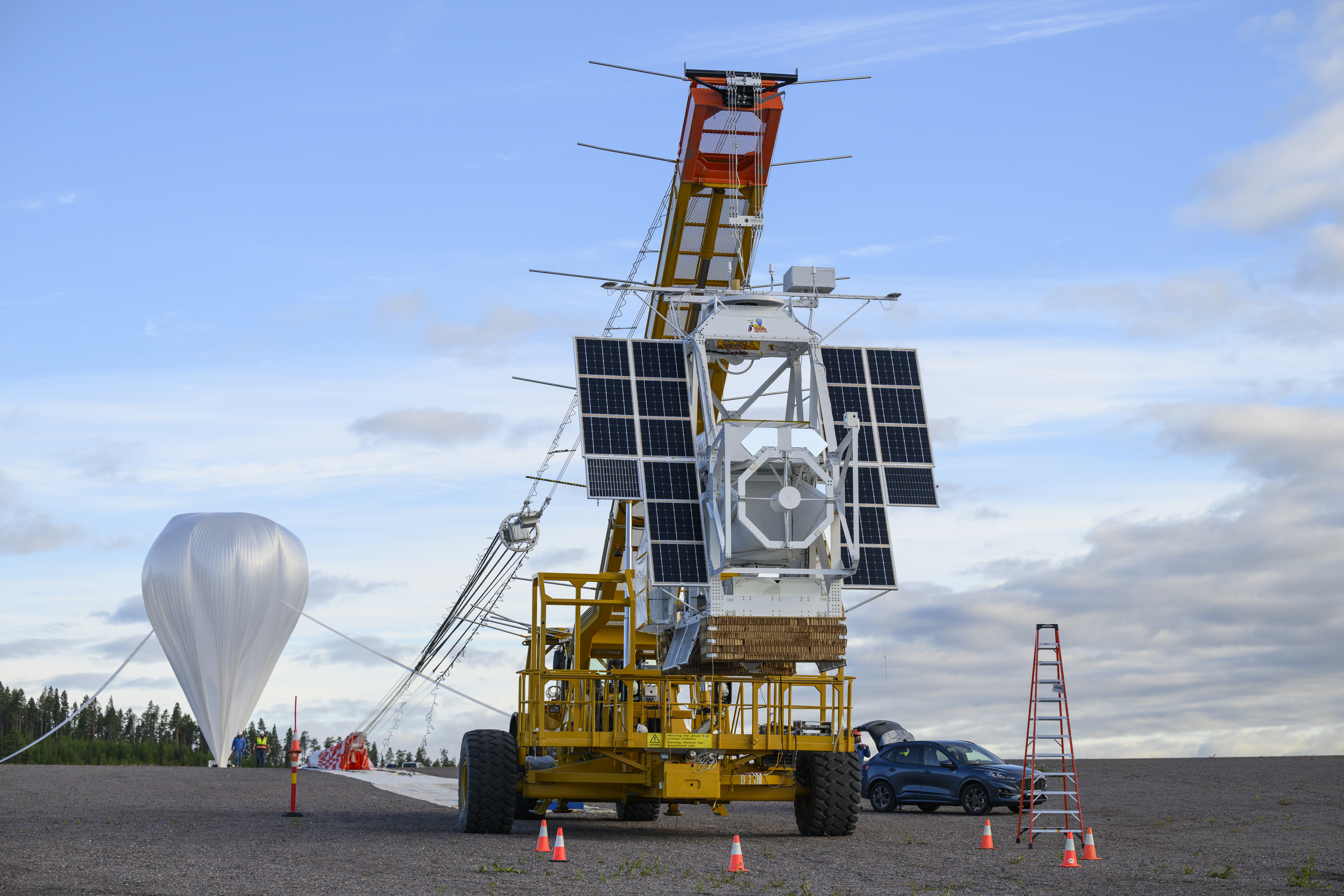}
   \end{tabular}
   \end{center}
   \caption{\sunriseiii{} before launch in 2024} 
    \label{sunrise} 
   \end{figure}

After two successful science flights in 2009 and 2013  \cite[see][]{solanki10,solanki17}, and a failed attempt in 2022, 
the \sunrise{} balloon-borne solar observatory \cite[see][]{barthol11,gandorfer11,berkefeld11,martinezpillet11} successfully completed its third science flight in July 2024 \cite[]{lagg25}, again from Kiruna, Sweden to northern Canada (see Figure~\ref{sunrise} showing \textsc{Sunrise~iii} on the launchpad of Esrange Space Center).
Stratospheric balloon-borne telescopes have two fundamental advantages over
ground-based telescopes: they permit UV observations and they provide a seeing-free
image quality over the full field of view (FoV) for in principle long periods of time. However, pointing to the Sun
and tracking a feature on the solar surface is a challenging task, especially for a
telescope hanging under a balloon that is driven by stratospheric winds at an
altitude of 36 km. In addition to the apparent (diurnal and seasonal) motion of
the Sun, there are a number of oscillatory modes that may be induced by variable
winds in the stratosphere, and taken up by the balloon-gondola system
The resulting gondola pendulum motion of a few degrees must be reduced to a few milliarcseconds to provide diffraction-limited observations. 

The gondola pointing system for \sunriseiii{} has been built by the Johns Hopkins University Applied Physics Lab (see \cite{gondola25}) and provides the coarse image stabilisation to within a few arcseconds. 
The second (fine) level of image stabilisation to within five to ten milliarcseconds is performed inside the post focus unit by the combination of a correlating wavefront sensor (CWS) and a fast tip-tilt mirror which measure and correct the residual image jitter, respectively. Furthermore, the CWS is also able to sense coma and focus. This information is used for in-flight telescope alignment using the secondary mirror M2. 

The CWS  was completely redesigned for the \sunriseiii{} flight in 2022 for the following reasons:
\begin{itemize}
\item due to a completely new suite of science instruments for \sunriseiii{}, the CWS operates at 640\,nm instead of 500\,nm as for \sunriseiandii{}, 
which would have led to a decreased measurement accuracy as a result of the longer wavelength and the lower contrast of the solar granulation. Adding a fast correlation tracker to the CWS compensates the loss of accuracy, but required a more powerful realtime control computer. 
\item advances in computing technology, i.e. changing the realtime control computer, allow for a simplified electronics setup and lower latencies in the realtime control loop of the tip-tilt correction, leading to an improved system performance even with the slower (but necessary) SUNRISE III tip-tilt mount, i.e. for tip-tilt, the 0-db correction bandwidth increases from 90\,Hz to 130\,Hz. 
\item in 2022 the CWS flight hardware of \sunriseiandii{} would have been more than 15 years old. Problems with batteries buffering some BIOS settings in the CWS realtime-computer added to the decision for using a new control computer.
\end{itemize}
Compared to the 2022 flight attempt, only minor software updates were applied for the 2024 science flight.

Section \ref{specs} lists the basic requirements and systems specifications, 
section \ref{sec:opticaldesign} explains the optical design, section \ref{sec:wfc} deals with the wavefront
correction algorithms and software. The electronics and tip-tilt hardware are described in section
\ref{sec:electronics}, followed by a performance analysis and a short conclusion.

\section{System Requirements / Specifications} \label{specs}

Table \ref{spectable} lists the requirements for the basic parameters of the \ac{cws}.  
The main differences to \sunriseiandii{} are the  operating wavelength and input voltage, which required changes in the optics and electronics. 
\begin{table}[ht]
\caption{CWS System Requirements}
\label{spectable}
\begin{center}       
\begin{tabular}{lc} 
\rule[-1ex]{0pt}{3.5ex} Parameter & Requirement\\
\hline
\rule[-1ex]{0pt}{3.5ex} measured aberrations & tip-tilt, focus, coma \\
\rule[-1ex]{0pt}{3.5ex} operating wavelength & 640\,nm  \\
\rule[-1ex]{0pt}{3.5ex} uninterrupted locking duration & $>$\, 4\,hours \\
\rule[-1ex]{0pt}{3.5ex}   residual TT correction accuracy & $<5$ milliarcseconds RMS  \\
\rule[-1ex]{0pt}{3.5ex}   TT correction bandwidth (0db) & $>$\, 100\,Hz \\
\rule[-1ex]{0pt}{3.5ex}  autofocus accuracy & $< \lambda/100$ RMS on granulation \\
\rule[-1ex]{0pt}{3.5ex}  coma measurement accuracy & $< \lambda/100$ RMS on granulation \\
\rule[-1ex]{0pt}{3.5ex}  TT correction range on sky &  60 arcseconds\\
\rule[-1ex]{0pt}{3.5ex}  input voltage & 24V \\
\rule[-1ex]{0pt}{3.5ex}  allowed power consumption at input & 80\,W \\
\rule[-1ex]{0pt}{3.5ex}  allowed mass (excl. TT mirror) & max. 10\,kg \\
\rule[-1ex]{0pt}{3.5ex} temperature range op / non-op [$^{\circ}$C] & 0\dots 30 / -40\dots 40 \\ 
\rule[-1ex]{0pt}{3.5ex} pressure range & 3\dots1000 mbar \\
\hline
\end{tabular}
\end{center}
\end{table} 
Although these requirements are very similar to the ones of \sunriseiandii{}, numerous modifications and innovations were necessary to interface with the new gondola, and to benefit from the damping of the pendulum motion with the gondola roll wheel. In addition, the science focus for \sunriseiii{} changed to obtaining several hour-long time series, requiring optimizations to achieve an uninterrupted locking duration of $>$\,4 hours.

\section{Optical Design}\label{sec:opticaldesign}
\subsection{Telescope}
The core component of the \sunrise{} observatory is a 1\,m Gregory-type telescope \cite[]{barthol11}. The parabolic primary mirror M1 (f/2.42) provides a
full disk image in its focal plane. Here a field stop, which is also a heat rejection
device, reflects and absorbs 99\% of the sunlight. Only a field of view of 200 arcsec passes to
the elliptical secondary mirror M2 which delivers a f/24.2 focus via two folding mirrors to F2, where a field stop is placed. Here, also several calibration targets, as well as a shutter for dark exposures can be inserted into the beam.
An Offner relay creates a real image of 25\,mm diameter of the system pupil on a fast tip-tilt mirror for compensating image jitter \cite[see][for a detailed description]{lagg25}.
Taking into account the maximum stroke of the tip-tilt mirror of 60 arcsec per axis, the maximum unvignetted science field of view has an edge length of 80 arcsec.
The light distribution of \sunriseiii{} feeds three scientific instruments: 
\begin{itemize}
\item SUSI (\sunrise{} Spectropolarimeter and Imager), a full Stokes spectropolarimeter working from 309 to 417\,nm, with slitjaw imaging and phase diversity capability for spectrum/slit and image reconstruction to remove remaining static aberrations \cite[]{susi25},
\item \textsc{TuMag} (Tunable Magnetograph), an imaging full Stokes spectropolarimeter working at 517 and 525\,nm and phase diversity capability \cite[]{tumag25}, and 
\item SCIP (\sunrise{} Chromospheric Infrared spectro-Polarimeter), a full Stokes spectropolarimeter working at 854 and 770\,nm with slitjaw imaging \cite[]{scip25}.
\end{itemize}
The fourth channel feeds the CWS which is responsible for the high speed tip-tilt correction, slow autofocus and coma measurement (coma is corrected manually).

\subsection{CWS}
In contrast to \sunriseiandii{}, where the CWS had a six-element Shack-Hartmann wavefront sensor (WFS) for fast tip-tilt correction, slow autofocus and coma measurement, 
\sunriseiii{} was upgraded with a correlation tracker (CT) for the fast tip-tilt-correction. The WFS is now only used for the slow autofocus and coma measurement. 
For the fast tip-tilt-correction this has the advantage of an improved tip-tilt measurement accuracy by using diffraction limited sampling of the full aperture resolution and a much faster readout (7\,kHz instead of 1.7\,kHz) of the CT camera which results in an increased correction bandwidth (130\,Hz instead of 90\,Hz).

\begin{figure} [ht]
   \begin{center}
   \begin{tabular}{c} 
   \includegraphics[height=6cm]{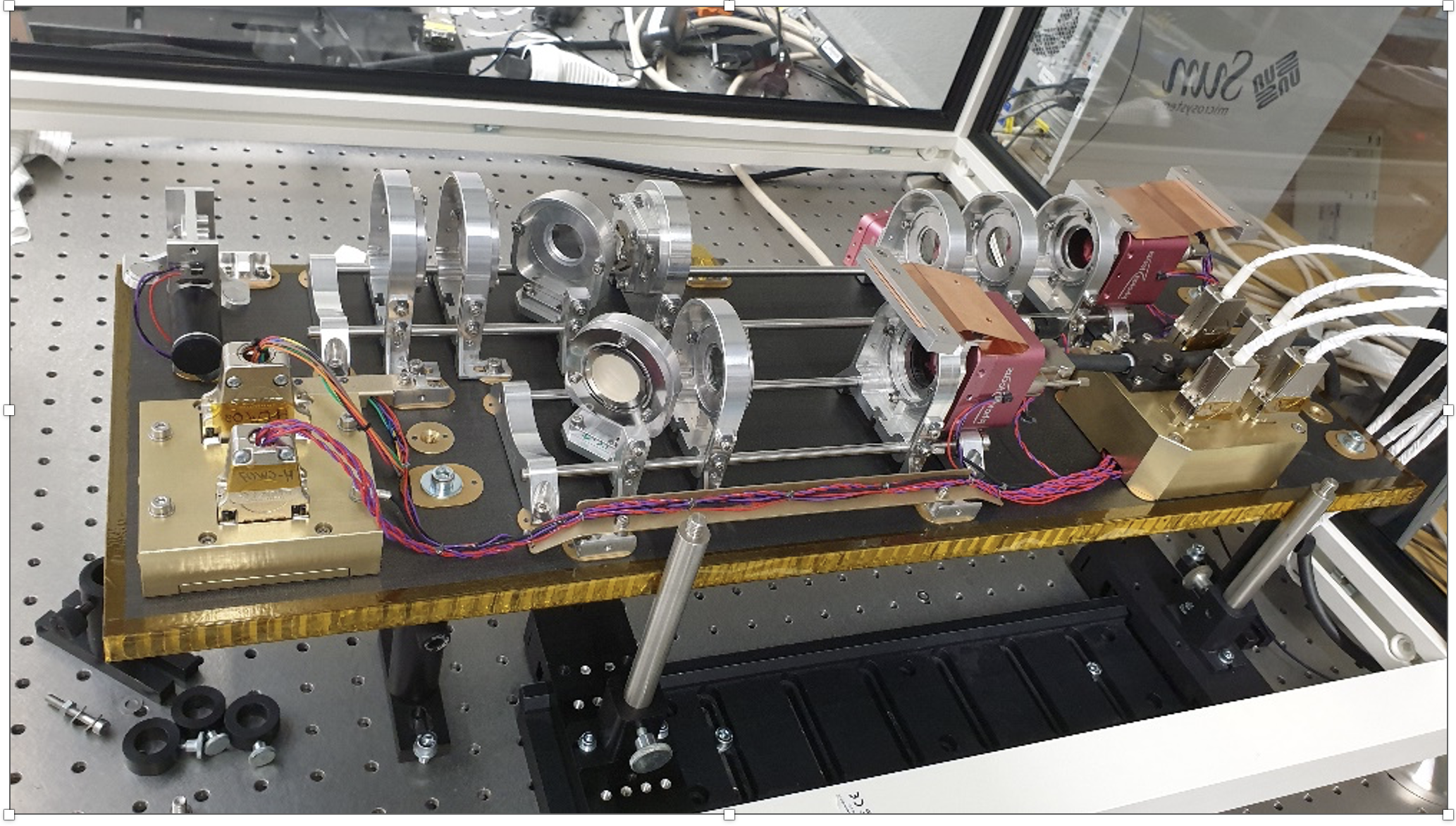}
   \end{tabular}
   \end{center}
   \caption{Optics (O-Unit) of the CWS. The light is coming from the left.} 
    \label{ounit} 
   \end{figure}

\begin{figure} [ht]
   \begin{center}
   \begin{tabular}{c} 
   \includegraphics[height=3cm]{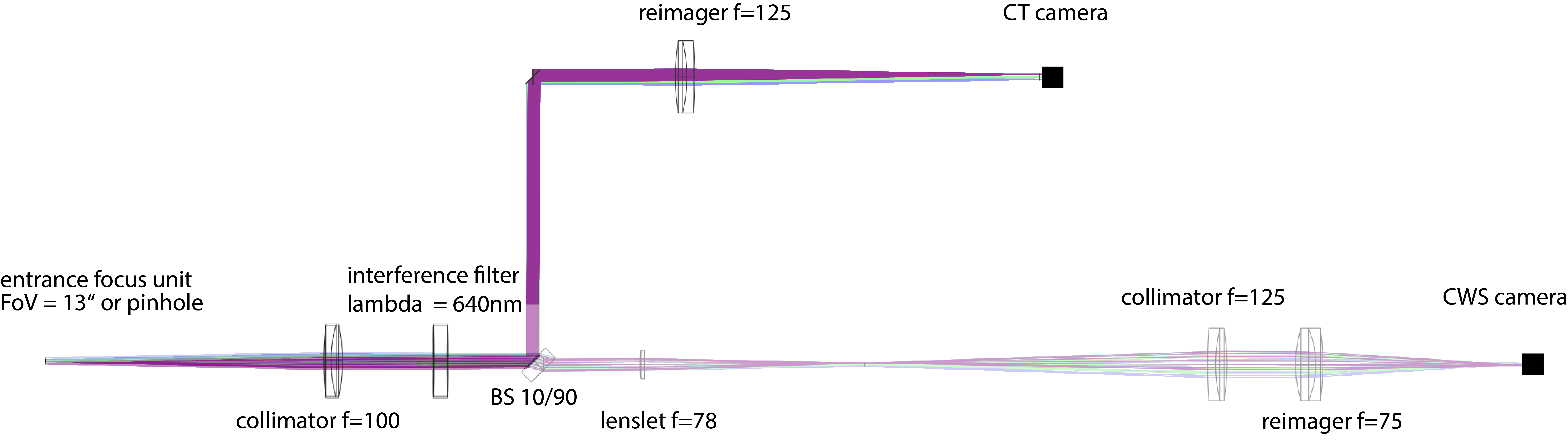}
   \end{tabular}
   \end{center}
   \caption{Optics design of the CWS. The light is coming from the left.} 
    \label{ct+cws-optics} 
   \end{figure}

Figures \ref{ounit} and \ref{ct+cws-optics} show the optics of the CWS, table \ref{ounittable} lists its parameters. 
The entrance focus unit of the CWS (the light enters from the left) includes a dark stop, a pinhole and a field stop which limits the field of view of the incoming f/24 beam to 13 arcsec on sky. The following collimator lens (f=100\,mm) collimates the beam. 
A bandpass filter with a central wavelength of 640\,nm and a bandpass of 10\,nm limits the photon flux and the wavelength range to minimise chromatic aberrations. 
Next, a beam splitter plate transmits 10\,\% of the light to the WFS and reflects 90\,\% to the CT. The splitting ratio reflects the much faster frame rate / lower exposure time 
of the CT as compared to the WFS. 

Hang tests in Göttingen in 2021 showed that the light throughput of the telescope and the transfer path was higher than estimated. A neutral density filter was added in the common CWS path to lower the light level by a factor of four (not shown in Figures \ref{ounit} and \ref{ct+cws-optics}).

For both channels, a pupil of diameter 4.1\,mm is created by the collimator lens with a focal length of  $f=100$\,mm. In the WFS channel, a 6-element lenslet ($f=78$\,mm) at the pupil position creates six subimages. The subsequent $\times \, 0.6$ magnifying stage ($f=125$\,mm + $f=75$\,mm) leads to the final 0.2\,arcsec/pix on the WFS camera. The six subapertures allow to measure coma, defocus and tip-tilt (the latter was only used for sun-pointing tests from ground). 
In the CT channel, the reimager lens ($f=125$\,mm) leads to a $1.25$ magnification. Both channels have diffraction-limited sampling (assuming the 1.22$\lambda/D$ definition) with respect to their respective effective aperture size.

\begin{table}[ht]
\caption{Parameters of the WFS and CT inside the optics unit of the CWS.} 
\label{ounittable}
\begin{center}       
\begin{tabular}{lcc} 
\hline
\rule[-1ex]{0pt}{3.5ex}   & WFS & CT  \\
\hline
\rule[-1ex]{0pt}{3.5ex}  \# of subapertures (total) & 6 & 1  \\
\rule[-1ex]{0pt}{3.5ex}  measured aberrations & tip-tilt, focus, coma & tip-tilt \\
\rule[-1ex]{0pt}{3.5ex}  subaperture size projected on M1 [m] & 0.33 &  1  \\
\rule[-1ex]{0pt}{3.5ex}  FoV on sky [arcsec] & 13 & 6.7  \\
\rule[-1ex]{0pt}{3.5ex}  pixel scale [arcsec/pix] & 0.2 & 0.07   \\
\rule[-1ex]{0pt}{3.5ex}  \# of pixels used for crosscorrelation & $64\times64$ & $96\times 96$   \\
\rule[-1ex]{0pt}{3.5ex}  camera (manufacturer = Photonfocus) & MV1-D1024E-80-G2 & MV1-D1024E-160-CL  \\
\rule[-1ex]{0pt}{3.5ex}  typical framerate [Hz]& 500 & 7000 \\
\hline
\end{tabular}
\end{center}
\end{table} 

\subsection{Tip-Tilt Mirror}
The tip-tilt mirror used for the fast image stabilisation is situated at a pupil image of 25\,mm in diameter. Its mirror blank has a diameter of 30\,mm. The mirror is driven by  a two-axis piezo ceramic actuator from Physik Instrumente
(S-330.8). This piezo drive fulfills all requirements for the \sunrise{} image stabilisation (range on sky $>$ 60 arcseconds and high dynamics, i.e., a resonance frequency of 700\,Hz (channel 1) and 1200\,Hz (channel 2) using the \sunriseiii{} tip-tilt mirror cell). The piezo actuators have a total tilt range of approx. 6 mrad (corresponding to $\pm 30$ arcsec on sky) using a voltage range of 7…93\,V . The minimal step size corresponds to  0.1\,$\mu$rad (0.001 arcsec on sky). 

Both, the design of the \sunriseiii{} tip-tilt mirror cell, and the mechanical holder of the piezo actuator inside the post-focus instrumentation platform had to be changed with respect to the \sunriseiandii{} designs. The new \sunriseiii{} tip-tilt mirror cell was designed by MPS with the main goal of maintaining the flatness of the  mirror surface up to a temperature of $100^{\circ}$ C under solar load. MPS built an additional mechanical adapter made of titanium with INVAR flex blades, a concept similar to the mirror interface used inside the image stabilisation of the High Resolution Telescope (HRT, see \cite{Gandorfer2018}) of the Polarimetric and Helioseismic Imager (PHI) \cite[]{Gandorfer2018,solanki20} on board Solar Orbiter \cite[]{mueller20}. This design preserves the optical surface quality of the tip-tilt mirror at the expense of a higher oscillating mass - thus reducing the eigenfrequency of the tip-tilt-system which leads to a lower tip-tilt correction bandwidth and ultimately lower tip-tilt correction quality. 
MPS also replaced the mechanical interface of the piezo actuators to the structure to fit into the spatial constraints given by the new optics layout.

\begin{figure} [ht]
   \begin{center}
   \begin{tabular}{c} 
   \includegraphics[height=9cm]{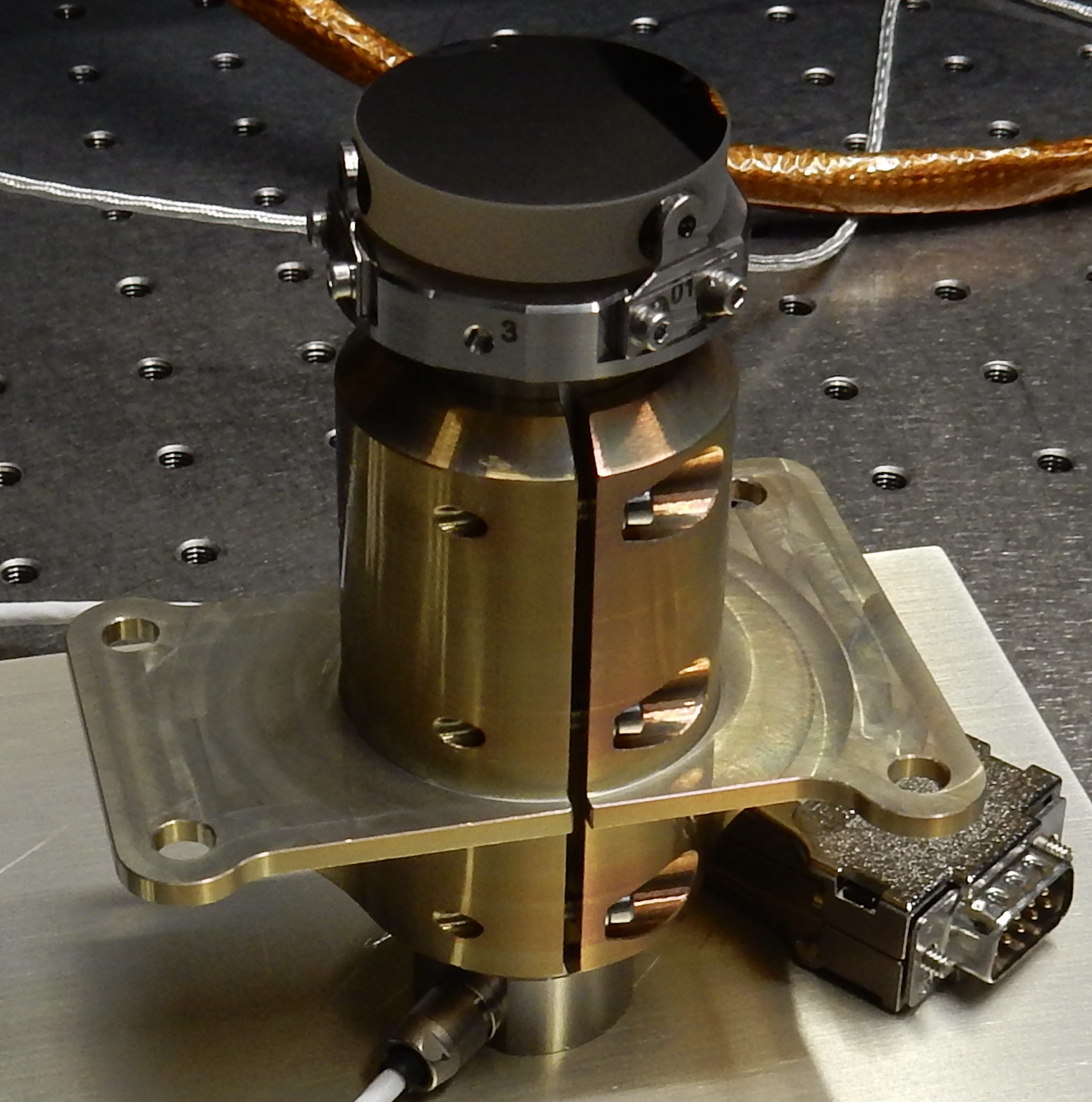}
   \end{tabular}
   \end{center}
   \caption{Picture of the tip-tilt mirror cell, the actuator, and its mount.} 
    \label{mirror} 
   \end{figure}

 Figure \ref{mirror} shows a picture of the tip-tilt mirror cell, the piezo actuator and its mechanical interface to the structure. 

\section{Wavefront Correction}\label{sec:wfc}
\subsection{Overview}
The main components of the wavefront correction system are the following
devices: the Shack-Hartmann wavefront sensor, the correlation tracker, each with a fast camera, 
the fast tip-tilt mirror (7000~Hz update rate), the slow secondary mirror M2 (0.1~Hz - 0.01~Hz update rate) and the control computer which connects the two sensors to the tip-tilt mirror and M2 (via the CWS communication software CW-COM (see Sec.~\ref{software}) and the ICU).

The telescope secondary mirror, M2, has high alignment tolerances with respect to M1, since a shift of M2 in axial (z) direction by, e.g., 16\,$\mu$m leads to a focus wavefront error of 100\,nm RMS in the science focus.
Therefore, M2 is mounted on a motorized {\it xyz} stage that is controlled by the CWS
to ensure proper alignment. Lateral ({\it xy}) misalignment causes coma and image
shift, while axial ({\it z}) misalignment causes spherical aberration but was found not be critical, 
and defocus.

\subsection{Wavefront Reconstruction}
Both the CT sensor and the Shack-Hartmann sensor use cross-correlation with a reference image for determining the image shift of the respective sub-aperture (the CT has only one sub-aperture which is the full aperture). 
Since the solar structures evolve, the cross correlation reference image has to be updated regularly, i.e. as fast as the solar structure evolves at a given spatial resolution. The reference images in both sensors were updated every 10 seconds.  

The data reduction of the sub-aperture pixel data is done in a very similar way as in
the solar AO system KAOS at the GREGOR solar telescope on Tenerife \cite[]{berkefeld2012}.

The wavefront reconstruction is also similar to the AO-system at the GREGOR solar telescope. 
The aberrations that need to be sensed determine the number of Shack-Hartmann sub-apertures: 
\begin{itemize}
\item Tip-tilt: needs to be corrected in closed loop, is sensed by the correlation tracker.
\item Defocus (caused by M2 or M3/M4 axial misalignment): needs to be corrected in a slow closed/automatic control loop to cope with focus drifts due to, e.g.,  temperature changes.
\item Coma (caused by M2 lateral misalignment): In principle, coma can be removed in closed loop by a 
lateral movement of M2. However, since a lateral movement of M2 produces 400 waves of tip-tilt for each wave of coma\footnote{One wave (640\,nm) rms of tip tilt 
aberration corresponds to 0.53 arcsec of image shift.}, a closed-loop coma-correction 
would have a negative impact on the tip-tilt performance. Instead, the coma is corrected manually prior to starting observation sequences. The initial alignment of the telescope ensures that 
only slight lateral M2 adjustments have to be made, so that no vignetting of the F2 field stop by the F1 field stop occurs. 
\item Spherical aberration (SA, caused by M2 axial misalignment): Although it is in principle possible to remove both, defocus and SA, simultaneously with M2 and M3/M4, an 
analysis of the optical errors showed that the amount of spherical aberration induced by a defocus of M2 is well within the range measurable by the phase diversity setups of the science instruments.
Therefore it was sufficient to statically align M2 and M3/M4 prior to the flight and correct the in-flight defocus with M2 only, and no sensing of SA was required.
\item Other higher order errors (such as triangular coma due to the M1 support): Interferometric measurements have shown that they have only minor contributions to the 
overall wavefront error. Furthermore, SUSI (UV) and \textsc{TuMag} (VIS) can apply phase diversity techniques to remove remaining small aberrations during post-facto processing. 
SCIP at its NIR wavelength range does not require this. 
\end{itemize}

The reconstruction matrix for the tip-tilt mirror is the
SVD-inverted measured interaction matrix between xy-shifts and the actuators.
The tip-tilt servo is a PID type (proportional + integral + differential), running at  6-7\,kHz loop frequency of the CT sensor.

Modal aliasing analysis shows that a six sub-aperture WFS is sufficient to sense defocus and coma (three modes). The condition number of the corresponding modes-to-shifts matrix which links the modal coefficients to the Shack-Hartmann WFS shifts is $<$ 5, i.e. well below 10 which is often used as a limit for numerical stability in AO wavefront reconstruction.

The corresponding reconstruction matrix has been pre-calculated via Zernike polynomials and scaled according to the WFS pixel scale. An additional conversion factor according to the M2 step size is applied inside the Instrument Control Unit. 
Since the focus drift due to
varying zenith angles occurs on timescales of minutes, M2 receives a
time-averaged signal. The averaging time is two seconds (1000 frames),  guaranteeing  a reduction of the measurement error below 0.05 rad RMS for defocus and coma, but also allowing a fast activation of the autofocus when starting observations.
Focus signals below a certain threshold (typically around  5\,$\rm \mu$m) were not applied to limit the number of M2-z movements, since they would cause an additional short pulse of lateral image motion.

\subsection{Measurement Accuracy}

Optics theory, and more specifically the equation given by \citet{michau}, show that the measurement accuracy of a correlating Shack-Hartmann sensor depends on the following
properties (which partially depend on each other): 
\begin{itemize}
\item Image contrast and its spatial power spectrum, i.e. the signal,
\item sub-aperture diameter,
\item observing / sensing wavelength,
\item pixel scale / sampling,
\item number of pixels per cross correlation field, and
\item noise (shot noise, camera readout noise and digitisation noise).
\end{itemize}

\begin{figure} [ht]
   \begin{center}
   \begin{tabular}{c} 
   \includegraphics[height=45mm]{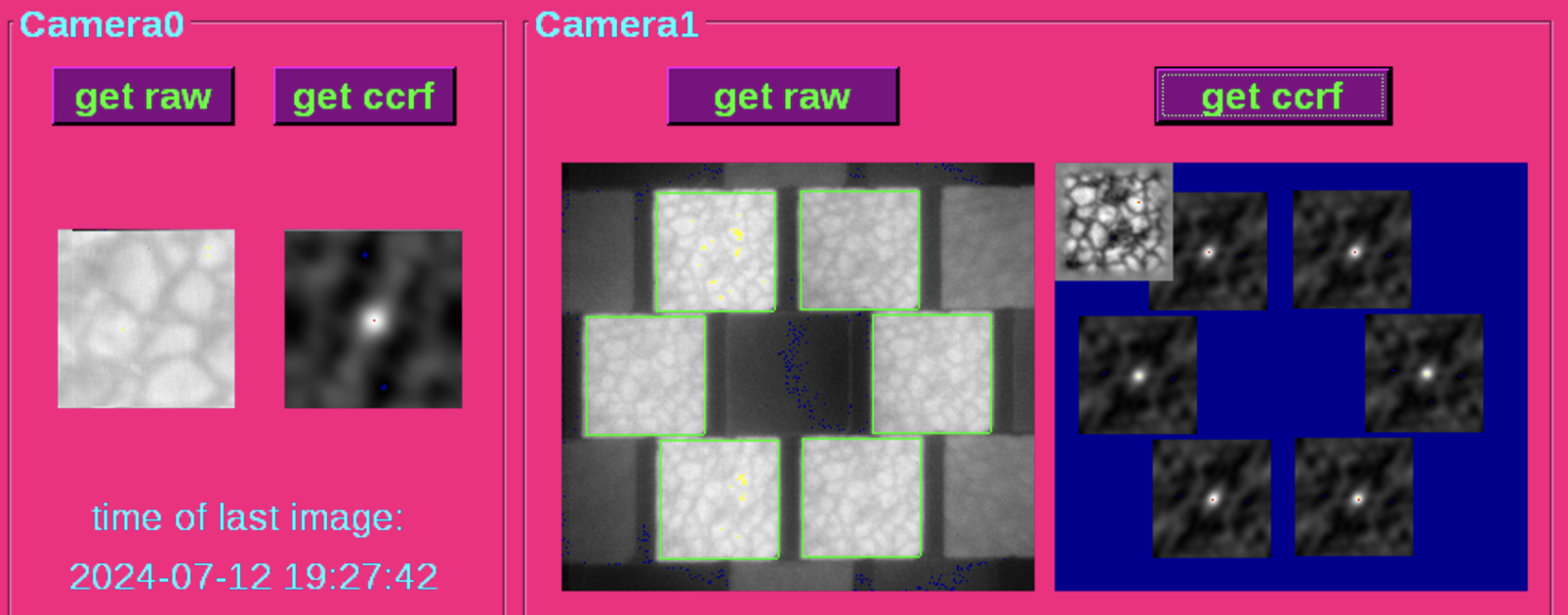}
   \end{tabular}
   \end{center}
   \caption{Screenshot of granulation showing the CT and WFS image plus their respective correlation function taken during the  \sunriseiii{}  flight.}
    \label{screenshot} 
   \end{figure}

The \sunrise{} CT-sensor works at 640~nm and uses the full 1\,m telescope aperture, 
producing a spatial resolution of 0.16 arcsec, using the 1.22\,$\lambda/D$ definition. The slight oversampling (0.07 arcsec/pixel) and
cross correlation field of 96$\,\times\,$96 pixels yield a field of view of 6.7 arcsec.  To
calculate the RMS intensity contrast as seen by the wavefront sensor, the original power
spectrum of the solar granulation has to be multiplied by the modulation
transfer function of the (sub-)aperture. This reduces the original granulation contrast of 11\,\% at 640\,nm, as measured by the Swedish Solar Telescope using Adaptive Optics and image reconstruction \cite[at 630\,nm, see ][]{scharmer10} 
to  8\,\% for the CT (theoretical contrast), and to 5\,\% for the WFS (theoretical contrast).
Image \ref{screenshot} is a live screenshot of granulation taken during the  \sunriseiii{} flight, showing both the CT and the WFS image plus their respective correlation function. The image contrast was 6.6\% (CT) and 4.9\% (WFS).

\citeauthor{michau}  estimated that the measurement noise variance 
$\sigma^2_{\rm snr, \lambda}$ (in $\lambda^2$) for a critically sampled, correlating
single aperture is 
\begin{equation}
 \sigma^2_{\rm snr, \lambda} = \frac{5\,m^2\sigma_{\rm noise}^2}{4\,n^2\sigma^2_{\rm signal}}, \label{eq:acc}
\end{equation}
where $m$ denotes the width of the autocorrelation function in pixels, $\sigma_{\rm noise}$ the noise in electrons standard deviation, $n$ the number of pixels
across the cross correlation field of view (96 for the CT) and $\sigma_{\rm signal}$ the
image contrast in electrons standard deviation.

The noise has three main contributors: for the photon noise
we assume a number of 160000~e$^-$ (photo electrons) per pixel (80\,\% full
well capacity), yielding 400~e$^-$ of shot noise.  The 8-bit digitization leads to a noise of
200000~e$^-$/256~$\approx$ ~800~e$^-$ peak-to-peak (LSB) which corresponds to about
(800/$\sqrt{3}$)~e$^-$~RMS = 462~e$^-$~RMS. The third contributor is the readout noise of 220~e$^-$ as specified by the camera manufacturer, resulting in a total noise of 650~e$^-$ standard deviation (assuming that the three noise sources are independent of each other).

For the signal we use 80\,\% of the full well capacity again and a contrast of 6.6\% (CT, corresponding to an average signal of ca 10500~e$^-$ / pixel) and 4.9\% (WFS, corresponding to a signal of 7800~e$^-$), respectively. 
Using Eq. (\ref{eq:acc}) and subsequent conversion from a wavefront error (in $\lambda$) to a pointing error (tip-tilt error, in arcsec), the RMS measurement error becomes
\begin{equation}
 \sigma_{\rm snr, \lambda} = 0.0007\cdot m{\rm\,\,   [\lambda], \,\,corresponding \,\,to\,\,\,}   \sigma_{\rm snr, tiptilt} \approx 0.00038\cdot m\,{\rm [arcsec]} 
\end{equation}
for the CT's 1\,m aperture and 
\begin{equation}
\sigma_{\rm snr, \lambda} = 0.0015\cdot m{\rm\,\, [\lambda]}
\end{equation}
for a WFS single 33\,cm subaperture.

For ground-based adaptive optics wavefront sensing on granulation, the correlation function is usually 4 pixels wide (for a typical subaperture size of 8-10\,cm and seeing in the same ballpark). 
However, in the case of granulation, the width $m$ of the correlation functions is 5-6 pixels for the \sunriseiii{} WFS and $\approx$12 pixels for the CT. 
The increased widths of the correlation functions for the \sunriseiii{} larger WFS and CT apertures resemble the decreasing solar signal at increasing spatial frequencies.  
When locking on sunspots the width of the correlation function can get even larger, however, the overall measurement error decreases due to the increased image contrast as compared to granulation.

The final CT measurement error for a single measurement when locking on granulation ($m=12$) becomes 
 \begin{equation}
 \sigma_{\rm snr, tiptilt} \approx 0.0045\,{\rm [arcsec]} .
\end{equation}
Due to the temporal averaging / oversampling of the CT control loop with respect to the correction bandwidth (7\,kHz as compared to 130\,Hz) the effective measurement error is significantly lower, but hard to quantify.

For the WFS (single subaperture), using $m=6$ results in 
 \begin{equation}
 \sigma_{\rm snr, \lambda} = 0.009{\rm\,\,   [\lambda]}.
\end{equation}
In theory, even without spatial (6 subapertures available, only 2 subapertures required for focus detection) or temporal averaging the accuracy of the focus measurement is already better than the focusing precision given by the M2 mechanical resolution (1\,$\rm \mu$m).  
Nevertheless, temporal averaging over 1000 frames (2 sec) plus a minimum threshold to avoid lateral image jitter by the focus mechanism was used for \sunriseiii{}.

\subsection{Real Time Control / Software} \label{software}
This section deals with the CWS flight control software running on an embedded Intel-based micro-computer. 
The conduction-cooled computer features a quad-core
Intel i7 processor, 32 GB RAM and connects to one camera using Ethernet
while the second camera is connected via a frame-grabber card running
the camera link protocol.
The CWS-software is divided into two parts: the first (CW-AO) handles the fast tip-tilt correction loop plus the slow autofocus correction/coma measurement. The second (CW-COM) handles the communication to the ICU (commanding and telemetry). This division allowed using  
a variant of the Adaptive Optics control software used at the GREGOR Solar Telescope on Tenerife  
\cite[]{berkefeld2012} as CW-AO. Equally important, the CW-COM software could be developed independently.
\subsubsection{CW-AO}
In order to achieve a high tip-tilt system bandwidth for the fast tip-tilt correction, the delays in the control
loop have to be as small as possible. Listed below are the delays of the
correction:
\begin {itemize}
\item 30 - 50~$\mu$s = half of the exposure time. The exposure time depends on the target (sunspot, quiet sun) and on the target position (disk center, close to the limb). 
\item 75~$\mu$s = readout time, according to the camera manufacturer
\item 75~$\mu$s = compute time, mostly for the shift determination of the $96\,\times\,96$ pixel correlation field. In order to achieve a low compute time, the data reduction is vectorised (auto-vectorisation by the compiler), the 
cross-correlation uses the FFT library {\it FFTW}, measured with the computer's high resolution timer
\item 50~$\mu$s = output + digital to analog (DA) conversion, given by the output data rate and the DA electronics specification  
\item 1100~$\mu$s = tip-tilt mirror settling time to 90\% including high voltage amplifier (measured time between input signal at the high voltage amplifier on one channel and the induced voltage at the piezo of the other channel)
\end{itemize}
An additional 70\,$\mu$s for the average hold time (limited sampling frequency, 1/7\,kHz/2) has to be added to the total delay of ca. 1.4\,ms between the occurrence of a disturbance and its correction. 
One CPU core is assigned exclusively to the fast tip-tilt control loop in order to minimise temporal jitter  to  1-2\,$\mu$s RMS, measured over a period of 8 seconds.

\subsubsection{CW-COM}
Running on the same computer as CW-AO, the CW-COM software handles the communication to the ICU  and (on ground) the CWS Electronic Ground Support Equipment (EGSE), 
freeing the realtime control software from these tasks.

CW-COM controls CW-AO by a state machine which handles special tasks required by the time-line driven operation
of the instruments. Amongst these the autolock
mode is the most important, which guarantees that lock for the focus
as well as for the image stabilisation are achieved automatically during
flight and are reestablished in case one or both of them are
lost. This is important for the above mentioned time-line driven
operation and for the case that stable telemetry to the \sunrise{}
gondola is lost during flight.

\section{Electronics}\label{sec:electronics}
\subsection{Overview}
\begin{figure} [ht]
   \begin{center}
   \begin{tabular}{c} 
   \includegraphics[height=6.9cm]{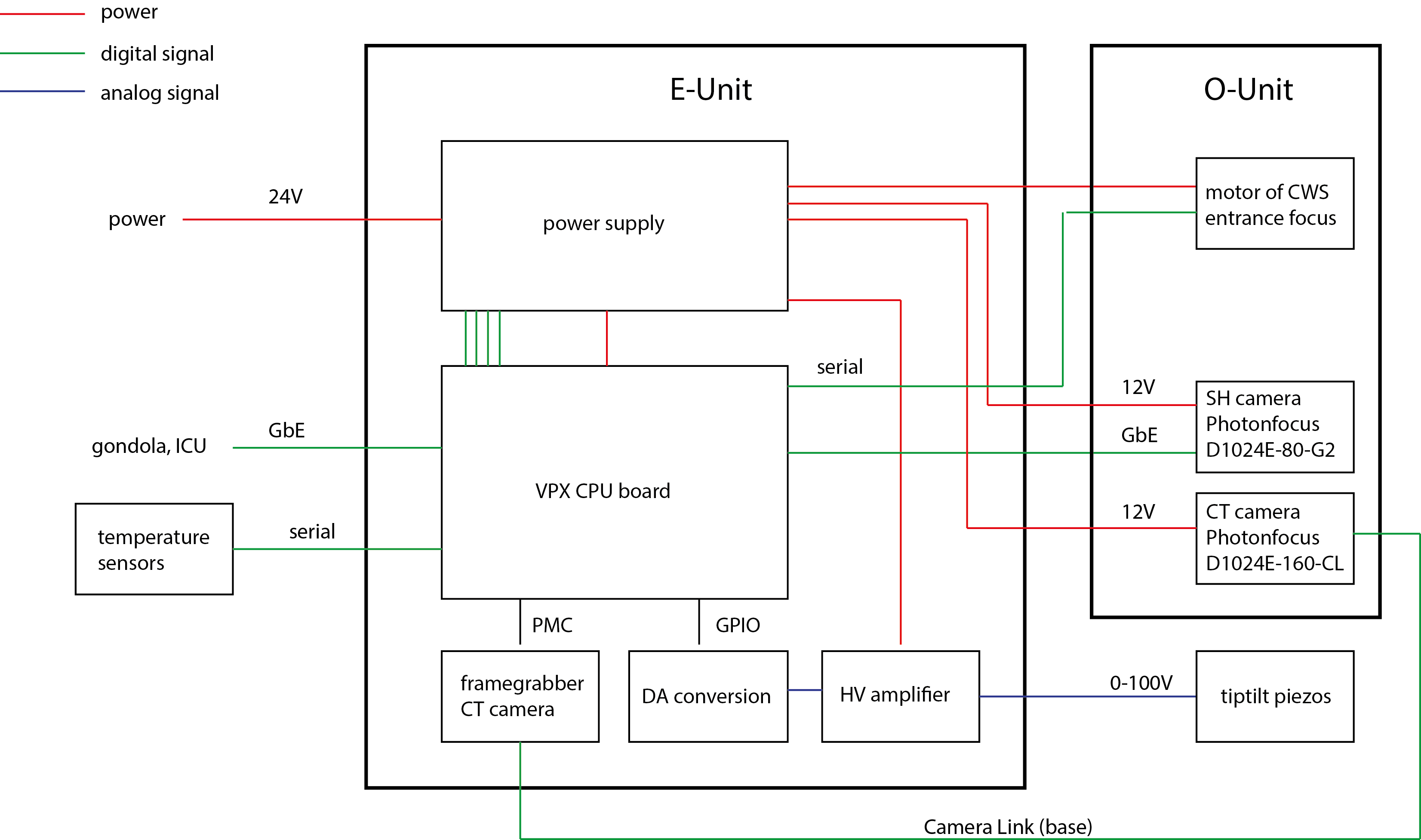}
   \end{tabular}
   \end{center}
   \caption{Electronics and harness layout.} 
    \label{electronics-setup} 
   \end{figure}

The CWS electronics for \sunriseiii{} differ significantly from the electronics for \sunriseiandii{} - the proximity electronics box (PEB) has disappeared and its contents have been split between the E-Unit and the O-Unit. As only the CWS entrance focus unit remains from the original three mechanisms, its motor controller could be integrated into the O-Unit and is directly connected to the E-Unit via RS422. The E-Unit now also contains the amplifier with 100\,V charge pump for the tip-tilt piezo drive and the power supply for the entire CWS electronics and the VPX6 computer board, which now also handles all communication tasks that were outsourced to a smaller computer board for \sunriseiandii{}. Furthermore, the system voltage changed from 24\,V DC to 12\,V DC, which allowed an isolated 24\,V to 24\,V DC/DC converter to be eliminated. To summarise, the electronics architecture has become simpler and more powerful (see schematics in Fig. \ref{electronics-setup}).

\subsection{Main Electronics Unit (E-Unit)}
\begin{figure} [ht]
   \begin{center}
   \begin{tabular}{c} 
   \includegraphics[height=6cm]{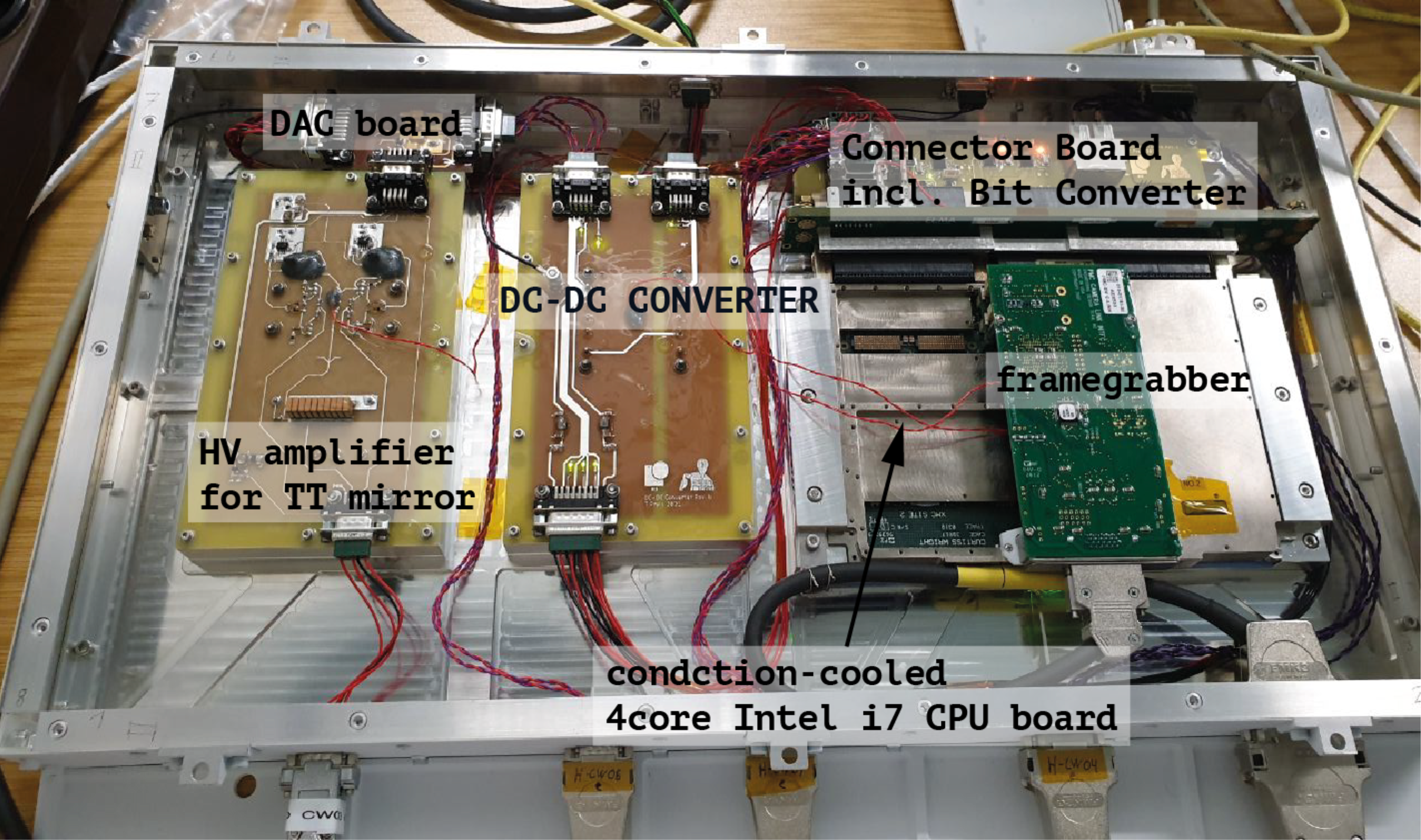}
   \end{tabular}
   \end{center}
   \caption{Opened E-unit with unmounted connectors: amplifier with DAC board (on the left), power supply (in the middle), VPX6 computer board with frame grabber (the green one) and the connector board at the back side.} 
    \label{e-unit} 
   \end{figure}

The main components of the E-unit are shown in Figure \ref{e-unit}:
The TT (tip-tilt) amplifier board has the identical circuit technology as in \sunriseiandii{}. Only the form factor has been adapted for the placement in the E-Unit. In addition, the amplifier has a digital-to-analog converter (DAC) board with two 16-bit AD5541 converters at the input. The power for all components of the CWS is provided from the DC/DC converter board that has several voltage levels with different requirements on current and/or ripple. Power management is much simpler as compared to \sunriseiandii{}. The VPX6 computer board only has to take care of switching the TT amplifier with its 100\,V charge pump on and off. All other components are supplied immediately as soon as the ICU supplies power to the CWS. The VPX6-1959 Single Board Computer with its Quad-Core i7-5850EQ processors at 2.7\,GHz, made by Curtiss Wright, handles all computational tasks: CWS correction control loop, all communication tasks via Ethernet, PCI Mezzanine Card (frame grabber PMC DV C-Link), serial interfaces and IOs, i.e., no additional computer is required. A parallel 7-bit real-time output drives the two 16-bit 5541 DA converters to the TT high voltage amplifiers, which reduced the output time from 119\,µs to 50\,µs  as compared to \sunriseiandii{}.

\subsection{Electronics in O-Unit}
The O-Unit contains two cameras, one motor controller and its DC-servo motor as main electronic components. 
The MV1-D1024E-3D02-160-G2 Ethernet camera is connected directly to the VPX6 computer board via Ethernet. It is supplied with 12\,V DC and its power requirement is less than 4.8\,W. The MV1-D1024E-160-CL Camera Link camera is connected to the frame grabber. It is also supplied with 12\,V and its power consumption is less than 3.2\,W. The motor controller in the box is a MCBL3002 P RS motion controller made by Faulhaber. The Faulhaber controller, which is also supplied with 12\,V, has an RS232 interface - an RS232-to-RS422 converter IC is included on the motor controller board in the box to ensure differential data transmission to the RS422 interface of the VPX6 computer board. The controlled 2036U024B K1155 brushless DC servo switches the state of the CWS entrance focus unit between dark stop, pinhole and field stop. 

\subsection{Tip-Tilt Mirror}
For the tip-tilt correction a two-axis piezo ceramic actuator from Physik Instrumente is used
(S-330.8). This piezo drive fulfills the requirements for the \sunrise{} image stabilisation (high acceleration, fast response to set points and small increments). The piezo actuators have a total tilt range of approx. 10\,mrad @ -20…120\,V, of which the range 7…93\,V is used, corresponding to ca 6\,mrad. The minimal step size given by the 16\,bit digitization is 0.1\,$\mu$ rad. 

For all \sunrise{} balloon flights, an in-house development for the piezo amplifier (specifications see Table~\ref{amplifier}) was necessary because no suitable off-the-shelf component was available. The PICO IRF charge pump, which supplies the 100\,V for the amplifier, is supplied with 12\,V for \sunriseiii{} (a PICO IRA charge pump for 24\,V DC was used for \sunriseiandii{}). The AB class amplifier has two channels, each with a complementary pair of mosfet transistors. The form factor had to be adapted for installation in the e-unit. 

\begin{table}[ht]
\caption{Amplifier Characteristics.} 
\label{amplifier}
\begin{center}       
\begin{tabular}{lc} 
\hline 
\rule[-1ex]{0pt}{3.5ex} Channel 1 & 7 V...93 V \\
\hline 
\rule[-1ex]{0pt}{3.5ex} Channel 2 & 7 V...93 V \\
\hline 
\rule[-1ex]{0pt}{3.5ex} Channel 3 & 100 V fixed \\
\hline 
\rule[-1ex]{0pt}{3.5ex} IRF charge pump supply & 8 V...24 V → 100 V (max. 10 W) \\
\hline 
\rule[-1ex]{0pt}{3.5ex} Push-pull power stage & max. 2 * 5 W @ 60 Hz full scale driven \\
\hline 
\rule[-1ex]{0pt}{3.5ex} Pre-amplifier & 0 V..5 V → 0 V..100 V \\
\hline 
\rule[-1ex]{0pt}{3.5ex} DA converter interface & SPI to 5x7-bit parallel (via micros) \\
\hline 
\rule[-1ex]{0pt}{3.5ex} DA converter update max. frequency & 20 kHz \\
\hline 
\rule[-1ex]{0pt}{3.5ex} DA converter resolution & 16 bit \\
\hline 
\rule[-1ex]{0pt}{3.5ex} Voltage scale & 97.6 V / 65535 = 1.489 mV / LSB \\
\hline 
\rule[-1ex]{0pt}{3.5ex} Dynamic amplifier range & 86 V \\
\hline 
\rule[-1ex]{0pt}{3.5ex} Tip-tilt range (on sky) & 60 arcsec \\
\hline 
\rule[-1ex]{0pt}{3.5ex} Tip-tilt resolution (on sky) & 0.001 arcsec \\
\hline 
\end{tabular}
\end{center}
\end{table}

\subsection{Thermal and Environmental Design}

At flight altitude the atmospheric pressure is about 5\,mbar. To dissipate the
power effectively by thermal radiation, the cooling elements have an allocated
radiation area, and the cooling design has to ensure that the heat is transmitted to that area. 
 A detailed thermal analysis of \sunriseiandii{} is described in 
\cite{gonzalezbarcena2022}.

For the CWS electronics unit we decided to work without a pressurised
box, and to dissipate the power only by conduction and radiation. All
electronics were designed for low power consumption and small power density
to decrease or even avoid any hot spots. The advantages of a conduction cooled
electronics unit in contrast to a pressurised box are reduced weight and a
rather simple temperature control with heaters and radiation coolers. Furthermore,
pressurised box and electronics designed to work under such pressure
pose the risk of failure due to loss of air pressure. To protect all components
from environmental influences, we applied an electrical isolation and hydrophobic
coating to all boards. The cooling elements had a white coating with high emissivity and low absorption (MAP SG121FD).
The power consumption of the CWS electronics unit in the operating mode is about 70\,W, of which 56\,W are used by the real time VPX6 computer board doing the wavefront correction. 

\begin{figure} [ht]
   \begin{center}
   \begin{tabular}{c} 
   \includegraphics[height=7cm]{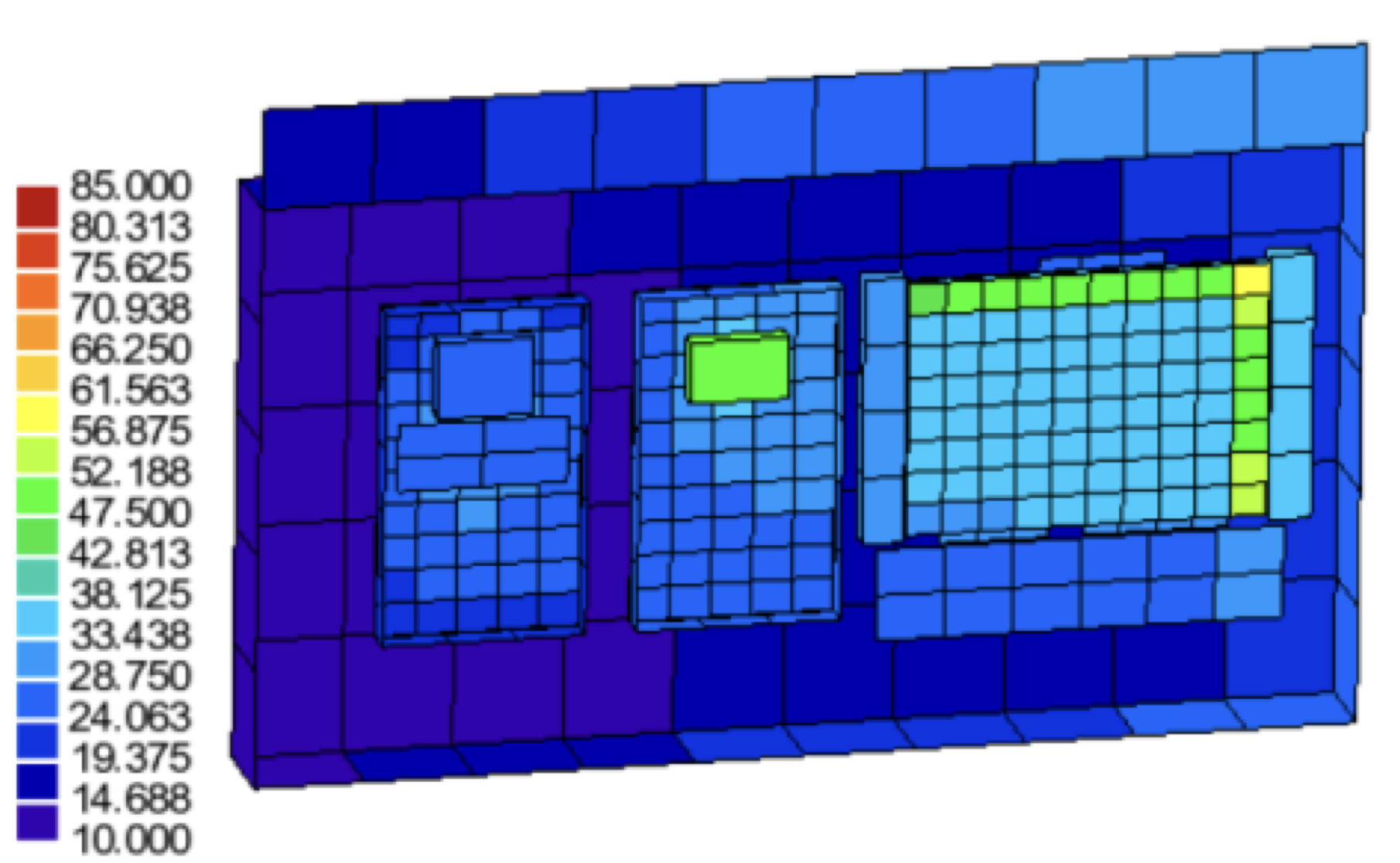}
   \end{tabular}
   \end{center}
   \caption{simulated temperature distribution (in deg C) of the E-Unit under flight conditions and in operation. 
   From left to right: amplifier for the tip-tilt piezo drive, the system power supply and the VPX6 computer board. 
   All temperatures are inside the operational limits.} 
    \label{temp} 
   \end{figure}

In order to check the thermal behaviour especially of the E-Unit which has the highest power consumption, a thermal FEM model of the CWS has been created (as an example, see the temperature distribution of the E-Unit in Fig. \ref{temp}).
The modeling results show that every part is inside the expected operational temperatures. 
This was confirmed by a four-day thermal vacuum test at 5\,mbar pressure which showed full functionality between $-30$ and $+40^\circ$C ambient temperature.
During the short flight in 2022 all CWS temperature sensors reported temperatures well inside the operational regime, both during the ascent phase and at flight altitude. 

\section{Performance}

\begin{figure} [ht]
   \begin{center}
   \begin{tabular}{c} 
   \includegraphics[height=7.3cm]{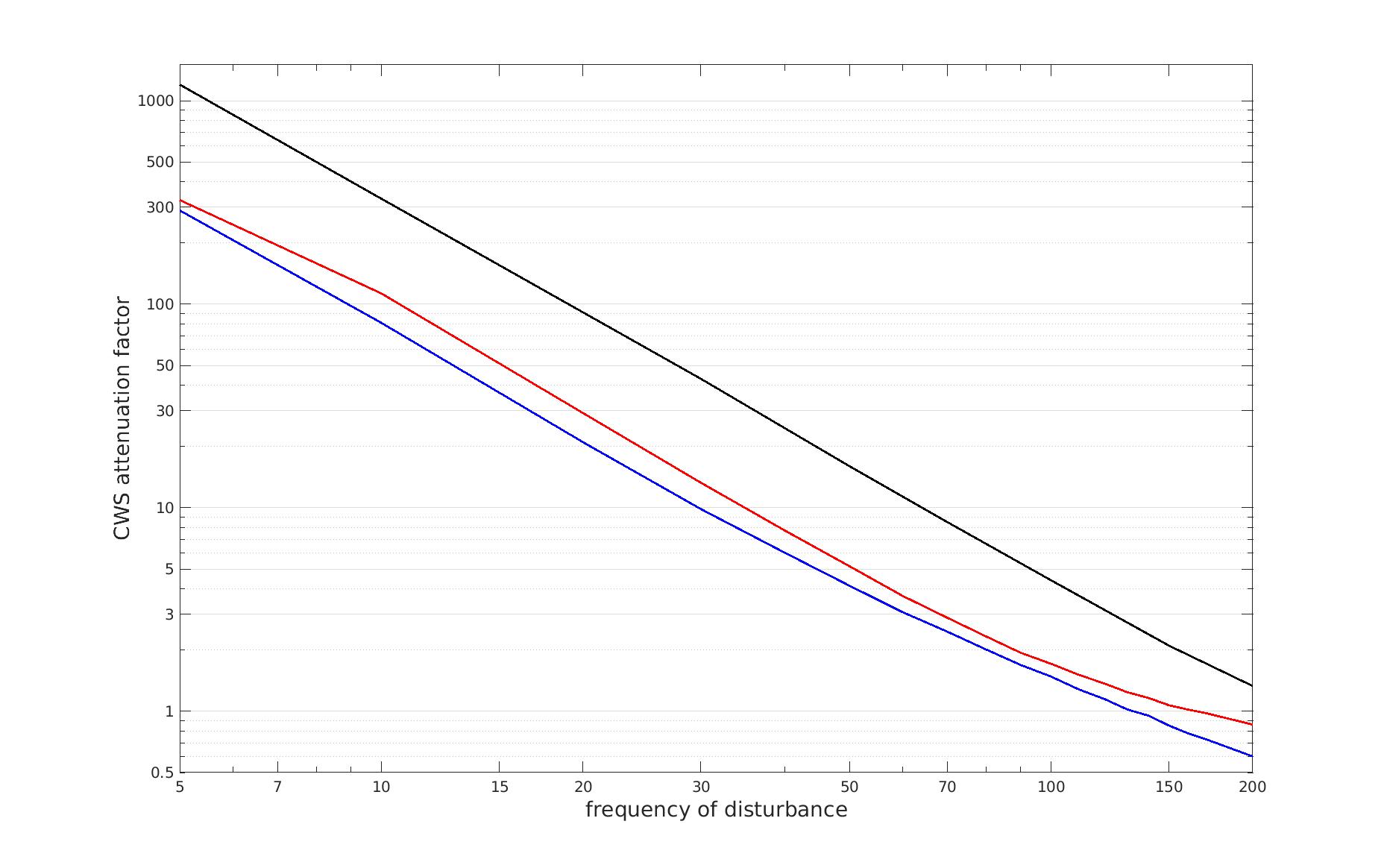}
   \end{tabular}
   \end{center}
   \caption{CWS \sunriseiii{} attenuation factor of the tip-tilt correction as a function of the frequency of the disturbance. \textsl{Black}: tip-tilt mirror mount used in \sunriseiandii{} (both axes), \textsl{blue}: flight model used in \sunriseiii{}, axis 1, \textsl{red}: flight model used in \sunriseiii{}, axis 2.} 
    \label{attenuation} 
   \end{figure}

The most important performance parameter of the \sunriseiii{} \ac{cws} is the attenuation factor for the tip-tilt correction. Residual image motions over a broad frequency range, from the gondola pendulum motion in the 0.1\,Hz range to high-speed vibrations up to 200\,Hz, are reduced by this factor.
Figure \ref{attenuation} shows the attenuation factors of the tip-tilt error as a function of the disturbance frequency measured during the integration of the system by applying a sine voltage to a second tip-tilt mirror which acted as an error source. The black curve denotes the performance of the old \sunriseiandii{} mirror mount, the red and blue curves resemble the \sunriseiii{} mirror mount.

The new \sunriseiii{} tip-tilt mount is optimised for a small (and constant) wavefront error under varying heat load, at the expense of a somewhat 
lower correction 0\,db bandwidth of 130\,Hz (axis 1) and 160\,Hz (axis 2), as compared to the 230\,Hz bandwidth when using the old \sunriseiandii{} tip-tilt  mount (both axes) in the \sunriseiii{} setup. As a result, the error attenuation when using the \sunriseiii{} mount 
is lower by a factor of 3 and 2.5, respectively, as compared to the \sunriseiandii{} mount used when in \sunriseiii{}. 
This behaviour is also reflected in the Proportional-Integral-Derivative (PID) servo parameters which had to be adapted to a slower servo response. The derivative (D)-part had to be turned down to almost zero. 

Nevertheless, the 130/160\,Hz tip-tilt correction bandwidth of  \sunriseiii{}  is significantly higher than the 90\,Hz of  \sunrisei{} and \sunriseii{}, and much higher than the initial (insufficient) 30\,Hz specification originating from \sunrisei{}. 

\begin{figure} [ht]
   \begin{center}
   \begin{tabular}{c} 
   \includegraphics[height=3.5cm]{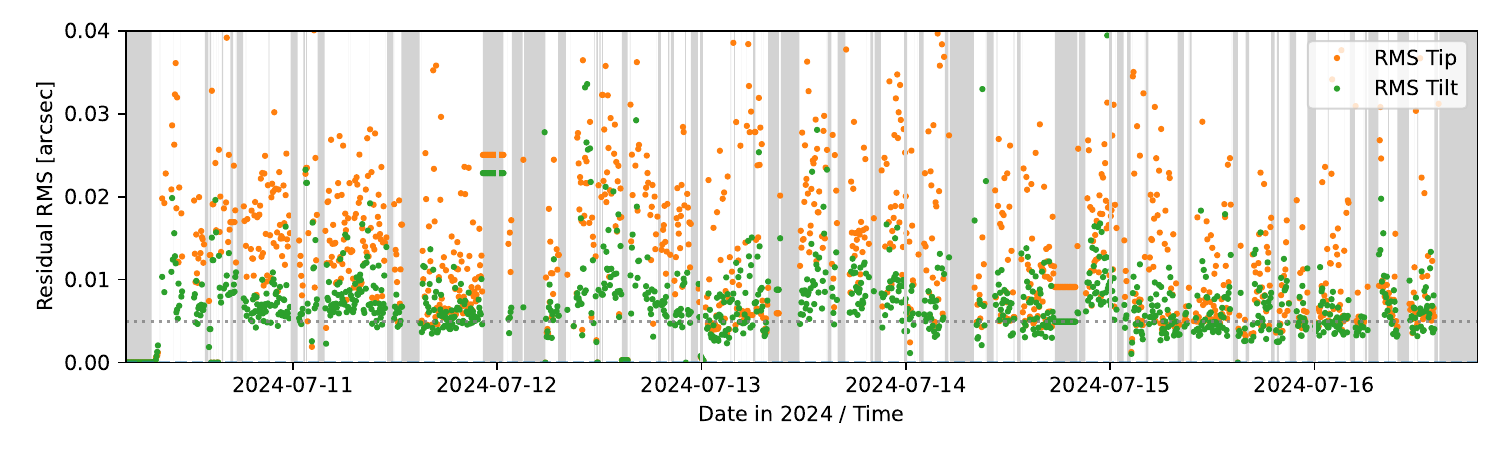}
   \end{tabular}
   \end{center}
   \caption{\sunriseiii{} residual (after correction) RMS image jitter as seen by the CWS as a function of time for tip (red) and tilt (green) axes, measured over 8 seconds intervals. The grey areas denote times when the CWS correction was off. The dotted horizontal line shows the 5 milliarcseconds specification.} 
\label{restt} 
\end{figure}

Figure \ref{restt} shows the in-flight residual RMS tip-tilt jitter at the science focus as measured by the CWS. It is obvious that the tilt axis performs much better than the tip axis. High speed telemetry showed the occurrence of hard external "shocks" which resulted in tip-tilt excursions of up to one arcsecond.  
The five milli-arcsecond specification is often met by the tilt-axis, but not by the tip-axis. 

During the last day of the flight, the gondola used softer control parameters which then transmitted less high frequency disturbances through the telescope.  As can be seen in Fig. \ref{restt}, this resulted in an overall better image stabilisation at the science focus. 

\section{Conclusion}
The image stabilisation system consisting of gondola and CWS played a crucial role for the success of the 2024 
\sunriseiii{} flight. During the six-day flight, more than 200\,TByte of high quality science data have been recorded. 

With a typical image stabilisation of $\approx$ 0.025'' RMS (total error of both axes), \sunriseiii{} has achieved - to our knowledge - the best image stabilization for balloon-borne mission to date.
Other missions achieved, for example, down to 0.05'' RMS over half an hour \cite[Balloon-borne Imaging Testbed,][]{romualdez} or 0.067'' RMS  \cite[Balloon Observation Platform for Planetary Science (BOPPS),][]{dankanich}. 
In future missions, gondola (coarse pointing system ) and fine pointing system should be optimised together as one system, not separately. 
In the case of \sunriseiii{} this was only done late in the flight, but resulting in residual jitter better than 0.01'' RMS for many (but not always contiguous) minutes. 
The possibility of stabilising a meter-class balloon-borne telescope down to 10 milliarcseconds  
opens the opportunity for future similar sized (and possibly even larger) balloon-borne telescopes, both for solar and night-time observations. 



\begin{acknowledgments}
We thank the Max Planck Institut für Sonnensystemforschung for their financial support.  A special thanks goes to the employees of CSBF and Esrange / SSC for their support and hospitality during the launch campaigns. We thank  our  workshops for their excellent work and F. Krämer for support during the flight.
\end{acknowledgments}

\begin{fundinginformation}
\sunriseiii{} is supported by funding from the Max-Planck-Förderstiftung (Max Planck Foundation), NASA under Grant \#80NSSC18K0934 and \#80NSSC24M0024 (``Heliophysics Low Cost Access to Space' program),  and the ISAS/JAXA Small Mission-of-Opportunity program and JSPS KAKENHI JP18H05234/JP23H01220. 
This research has received financial support from the European Union’s Horizon 2020 research and innovation program under grant agreement No. 824135 (SOLARNET) and No. 101097844 (WINSUN) from the European Resaerch Council (ERC). It has also been funded by the Deutsches Zentrum für Luft- und Raumfahrt e.V. (DLR, grant no. 50 OO 1608).
The Spanish contributions have been funded by the Spanish MCIN/AEI under projects RTI2018-096886-B-C5, and PID2021-125325OB-C5, and from ``Center of Excellence Severo Ochoa'' awards to IAA-CSIC (SEV-2017-0709, CEX2021-001131-S), all co-funded by European REDEF funds, ``A way of making Europe''. 

D.~Orozco~Suárez acknowledges financial support from a {\em Ram\'on y Cajal} fellowship.
The research activities and the flight operation of  M.~Kubo were supported from the JSPS KAKENHI grants No. 23KJ0299, No. 24K07105, No. 23K13152, and No. 21K13972, respectively.
\end{fundinginformation}


\bibliographystyle{spr-mp-sola}
\bibliography{main} 




\begin{acronym}
\setlength{\itemsep}{-4ex}

\acro{alma}[ALMA]{Atacama Large Millimeter/sub-millimeter Array}
\acro{amhd}[AMHD]{{Analog, Motor, and Heaters Drivers}}
\acro{ao}[AO]{adaptive optics}
\acro{bbso}[BBSO]{Big Bear Solar Observatory} 
\acro{bpo}[BPO]{Balloon Program Office}
\acro{cfrp}[CFRP]{carbon fiber-reinforced polymers}
\acro{chase}[CHASE]{Chinese \halpha{} Solar Explorer}
\acro{cmos}[CMOS]{complementary metal oxide semiconductor}
\acro{cnoc}[CNOC]{cold non-operational case}
\acro{coc}[COC]{cold operational case}
\acro{coi}[CoI]{Co-Investigator}
\acro{csbf}[CSBF]{Columbia Scientific Ballooning Facility}
\acro{ct}[CT]{correlation tracker}
\acro{cws}[CWS]{Correlating Wavefront Sensor}
\acro{dkist}[DKIST]{Daniel K. Inouye Solar Telescope}
\acro{docdb}[S3-DocDB]{\sunriseiii{} Document Database}
\acro{dss}[DSS]{Data Storage System}
\acro{dst}[DST]{Dunn Solar Telescope}
\acro{egse}[EGSE]{electrical ground support equipment}
\acro{elink}[E-Link]{Esrange Airborne Data Link}
\acro{eoc}[EOC]{Esrange Operations Center}
\acro{erack}[E-rack]{electronic rack}
\acro{esrange}[Esrange]{Esrange Space Center}
\acro{evtm}[EVTM]{Ethernet Via Telemetry}
\acro{fits}[FITS]{Flexible Image Transport System}
\acro{fov}[FoV]{field-of-view}
\acro{fps}[fps]{frames per second}
\acro{fram}[FRAM]{ferroelectric random access memory}
\acro{frr}[FRR]{Flight Readiness Review}
\acro{fsm}[FSM]{finite state machine}
\acro{fwhm}[FWHM]{full width at half maximum}
\acro{goc}[GOC]{Göttingen Operations Center}
\acro{gps}[GPS]{Global Positioning System}
\acro{gregor}[GREGOR]{GREGOR}
\acro{gst}[GST]{Goode Solar Telescope}
\acro{gusto}[GUSTO]{Galactic/Extragalactic ULDB Spectroscopic Terahertz Observatory}
\acro{hmi}[HMI]{Helioseismic Magnetic Imager}
\acro{hnoc}[HNOC]{hot non-operational case}
\acro{hoc}[HOC]{hot operational case}
\acro{hrw}[HRW]{Heat Rejection Wedge}
\acro{hvps}[HVPS]{high voltage power supply}
\acro{iaa}[IAA]{Instituto de Astrofísica de Andalucía}
\acro{iac}[IAC]{Instituto de Astrofísica de Canarias}
\acro{ics}[ICS]{Instrument Control System}
\acro{icu}[ICU]{Instrument Control Unit}
\acro{ihop}[IHOP]{IRIS/\hinode{} Operations Plan}
\acro{imax}[\textsc{IMaX}]{Imaging Magnetograph eXperiment}
\acro{imtek}[IMTEK]{Department of Microsystems Engineering}
\acro{imu}[IMU]{Inertial Measurement Unit}
\acro{inta}[INTA]{Instituto Nacional de Técnica Aeroespacial}
\acro{iridium}[Iridium Pilot]{Iridium Pilot\textsuperscript{\textregistered}}
\acro{iris1}[IRIS-1]{Image Recording Instrument for \sunriseii{}}
\acro{iris2}[IRIS-2]{Image Recording Instrument for \sunriseiii{}}
\acro{iris}[IRIS]{Interface Region Imaging Spectrograph}
\acro{ir}[IR]{infrared}
\acro{islid}[\textsc{ISLiD}]{Image Stabilization and Light Distribution unit}
\acro{jaxa}[JAXA]{Japan Aerospace Exploration Agency}
\acro{apl}[JHUAPL]{Johns Hopkins University Applied Physics Laboratory}
\acro{kbsi}[KBSI]{Korea Basic Science Institute}
\acro{kis}[KIS]{Institut für Sonnenphysik}
\acro{lcvr}[LCVR]{liquid crystal variable retarder}
\acro{ldb}[LDB]{long-distance balloon}
\acro{ld}[LD]{launch date}
\acro{led}[LED]{light-emitting diode}
\acro{lldpe}[LLDPE]{linear low-density polyethylene}
\acro{los}[LoS]{line-of-sight}
\acro{lte}[LTE]{local thermodynamic equilibrium}
\acro{mast}[MAST]{Multi-Application Solar Telescope}
\acro{mhd}[MHD]{magneto-hydrodynamic}
\acro{MHD}[MHD]{Magneto-Hydrodynamic}
\acro{mit}[MIT]{Massachusetts Institute of Technology}
\acro{mli}[MIL]{multilayer insulation}
\acro{mpae}[MPAe]{Max-Planck-Institut für Aeronomie}
\acro{mpg}[MPG]{Max-Planck-Gesellschaft}
\acro{mps}[MPS]{Max Planck Institute for Solar System Research}
\acro{mrr}[MRR]{Mission Readiness Review}
\acro{mtc}[MTC]{Main Telescope Controller}
\acro{muramce}[MURaM-CE]{MURaM Chromospheric Extension}
\acro{muram}[MURaM]{MPS/University of Chicago Radiative MHD}
\acro{naoj}[NAOJ]{National Astronomical Observatory of Japan}
\acro{nasa}[NASA]{National Aeronautics and Space Administration}
\acro{nic}[NIC]{network interface controller}
\acro{nir}[near-IR]{near-infrared}
\acro{nlte}[non-LTE]{non-local thermodynamic equilibrium}
\acro{nuv}[near-UV]{near-ultraviolet}
\acro{nvst}[NVST]{New Vacuum Solar Telescope}
\acro{occ}[OCC]{Operations Control Center}
\acro{olr}[OLR]{outgoing longwave radiation}
\acro{pbs}[PBS]{polarising beam-splitter}
\acro{pcb}[PCB]{printed circuit board}
\acro{pcu}[PCU]{Power Converter Unit}
\acro{pdu}[PDU]{Power Distribution Unit}
\acro{pfi}[PFI]{Post Focus Instrumentation platform}
\acro{phi}[PHI]{Polarimetric and Helioseismic Imager}
\acro{pi}[PI]{Principal Investigator}
\acro{pmu}[PMU]{Polarization Modulation Unit}
\acro{poe}[PoE]{Power over Ethernet}
\acro{ppd}[PPD]{PFI Power Distribution Unit}
\acro{psf}[PSF]{point spread function}
\acro{ps}[PS]{Pointing System}
\acro{hk}[HK]{housekeeping}
\acro{ramon}[RAMON]{RAdiation MONitor}
\acro{rms}[rms]{root-mean-square}
\acro{rtc}[RTC]{real-time clock}
\acro{s2n}[S/N]{signal-to-noise}
\acro{s3pc}[S$^3$PC]{Spanish Space Solar Physics Consortium}
\acro{scip}[\textsc{SCIP}]{\sunrise{} Chromospheric Infrared spectro-Polarimeter}
\acro{sdo}[SDO]{Solar Dynamics Observatory}
\acro{sgt}[SGT]{Sun Guider Telescope}
\acro{sipm}[SiPM]{Silicon PhotoMultiplier}
\acro{sip}[SIP]{Support Instrumentation Package}
\acro{smart}[SMART]{Solar Magnetic Activity Research Telescope}
\acro{smm}[SMM]{scan mirror mechanism}
\acro{solo}[SolO]{Solar Orbiter}
\acro{sophi}[SO/PHI]{Solar Orbiter Polarimetric and Helioseismic Imager}
\acro{sotnfi}[NFI]{Narrowband Filter Imager}
\acro{sotsp}[SP]{Spectro-Polarimeter}
\acro{sot}[SOT]{Solar Optical Telescope}
\acro{ssct}[SSCT]{Science/Support Compatibility Test}    
\acro{ssc}[SSC]{Swedish Space Corporation}
\acro{ssd}[SSD]{solid state disk}
\acro{ssh}[SSH]{Secure Shell}
\acro{ssm}[SSM]{second surface mirror}
\acro{sst}[SST]{Swedish Solar Telescope}
\acro{sswg}[SSWG]{\sunrise{} Science Working Group}
\acro{ss}[SS]{Science Stack}
\acro{starlink}[Starlink]{Starlink (a division of SpaceX)}
\acro{sufi}[\textsc{SuFI}]{\sunrise{} Filtergraph and Imager}
\acro{sunaps}[\textsc{SunAPS}]{\textit{\sunriseiii{} science planning And Pointing System}}
\acro{sunrise}[\textsc{Sunrise}]{{\textbf{S}olar \textbf{U}V to \textbf{N}ear-infra\textbf{R}ed \textbf{I}maging and \textbf{S}pectropolarimetric \textbf{E}xploration}}
\acro{susi}[\textsc{SUSI}]{\sunrise{} Spectropolarimeter and Imager}
\acro{tc}[TC]{Telecommand}
\acro{tdrss}[TDRSS]{Tracking \& Data Relay Satellite System}
\acro{tmtc}[TM/TC]{Telemetry and Telecommand}
\acro{tm}[TM]{Telemetry}
\acro{tumag}[\textsc{TuMag}]{Tunable Magnetograph}
\acro{tv}[TV]{thermal-vacuum}
\acro{unival}[UV]{Universitat de València}
\acro{upm}[UPM]{Universidad Politécnica de Madrid}
\acro{usaf}[USAF]{U.S. Air Force}
\acro{uv}[UV]{ultraviolet}
\acro{vda}[VDA]{vapor deposited aluminum}
\acro{vpn}[VPN]{virtual private network}
\acro{vtt}[VTT]{German Vacuum Tower Telescope} 
\acro{wfs}[WFS]{wavefront sensor}

\end{acronym}

\end{document}